\documentclass[pra,fleqn,showpacs,twocolumn]{revtex4}

\usepackage{amsmath,graphicx}

\begin{document}

\title{Spin-orbit corrections of order $m\alpha^6$ to the fine structure of $(37,35)$ state in $^4\mbox{He}^+\bar{p}$ atom.}
\author{Vladimir I.~Korobov}
\affiliation{Joint Institute for Nuclear Research\\
141980, Dubna, Russia}
\email{korobov@theor.jinr.ru}
\author{Zhen-Xiang Zhong}
\affiliation{Wuhan Institute of Physics and Mathematics, CAS \\
430071, Wuhan, People's Republic of China}

\pacs{36.10.-k, 31.15.A-,31.30.J-}

\begin{abstract}
Precise numerical calculation of radiofrequency intervals between hyperfine sublevels of the $(37,35)$ state of the antiprotonic helium-4 atom is presented. Theoretical consideration includes the QED corrections of order $m\alpha^6$ to the electron spin-orbit interaction. The effective Hamiltonian is derived using the formalism of the nonrelativistic quantum electrodynamics (NRQED).
\end{abstract}

\maketitle

\section{Introduction}

The high precision spectroscopy measurement of the hyperfine structure of the antiprotonic helium has two-fold interest. First, it is expected that it may be a way to obtain improved value of the magnetic moment of an antiproton. The other point is that it can be a good benchmark for testing QED theoretical methods for the Coulomb three-body bound states to a high precision.

\begin{figure}[b]\label{fig:HFS}
\includegraphics[width=0.40\textwidth]{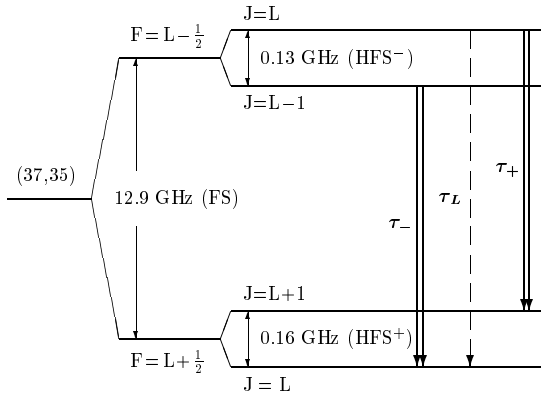}
\caption{Schematic diagram of hyperfine sublevels of the $(37,35)$ state
of $^4\mbox{He}^+\bar{p}$ atom.}
\end{figure}

At present several theoretical calculations for the hyperfine structure of the $(37,35)$ state of the $^4\mbox{He}^+\bar{p}$ atom have been performed \cite{PRA98,Kino01,JPB01,Kino03}. Since all the results were obtained within the frames of the same Breit-Pauli approximation, the major difference in the obtained data was either due to numerical inaccuracy of the nonrelativistic solution or due to a difference in the choice of the physical constants. Still they were in a good agreement (within the error bars of the theoretical approximation) with the first experimental observation of the 13 GHz intervals \cite{Wid02}.

Recently, the ASACUSA experiment has obtained new precise values for the two RF transitions of the $(37,35)$ state in the $^4\mbox{He}^+\bar{p}$ atom \cite{Wid08,PLB09} (the notation is shown on the schematic diagram of the $(37,35)$ state on Fig.~\ref{fig:HFS}):
\begin{equation}\label{exp}
\begin{array}{@{}l}
\tau_+ = 12\,896.641(63)\mbox{ MHz},
\\[2mm]
\tau_- = 12\,924.461(63)\mbox{ MHz}.
\end{array}
\end{equation}
The difference between $\tau_+$ and $\tau_-$ is nearly proportional to the antiprotonic magnetic moment and has a value
\begin{equation}\label{exp:2}
\Delta \tau = 27.825(33)\mbox{ MHz}.
\end{equation}

That should be compared with the theoretical prediction \cite{JPB01}:
\begin{equation}
\begin{array}{@{}l}
\tau_+ = 12\,896.35(69)\mbox{ MHz},
\\[2mm]
\tau_- = 12\,924.24(69)\mbox{ MHz}.
\end{array}
\end{equation}
It is seen that the experimental error is more than an order of magnitude smaller and transitions have some systematic shift toward larger values. The Breit-Pauli approximation used so far is limited by the uncertainty of order $\mathcal{O}(\alpha^2)$ and higher order corrections should be included into consideration to achieve the similar accuracy as in the experiment.

The effective Hamiltonian of the hyperfine interaction may be written (see details in \cite{JPB01})
\begin{equation}\label{eq:eff_H}
\begin{array}{l}
H^{\rm eff}\!=
   \underline{E_1\,(\mathbf{s}_e\!\cdot\!\mathbf{L})}+
   E_2\,(\mathbf{s}_{\bar{p}}\!\cdot\!\mathbf{L})+
   E_3\,(\mathbf{s}_e\!\cdot\!\mathbf{s}_{\bar{p}})
\\[3mm]\hspace{20mm}
   +E_4\,\bigl\{2L(L\!+\!1)(\mathbf{s}_e\!\cdot\!\mathbf{s}_{\bar{p}})
\\[3mm]\hspace{20mm}
     -3\bigl[(\mathbf{s}_{\bar{p}}\!\cdot\!\mathbf{L})(\mathbf{s}_e\!\cdot\!\mathbf{L})
     +(\mathbf{s}_e\!\cdot\!\mathbf{L})(\mathbf{s}_{\bar{p}}\!\cdot\!\mathbf{L})\bigr]
\bigr\}.
\end{array}
\end{equation}
To get improved values for the $\tau_+$ and $\tau_-$ transitions one needs to get contributions of the next to the leading order for the electron spin-orbit interaction coefficient $E_1$. It may be done within the framework of the NRQED formalism \cite{Kin96}. Some details of the derivation of the $m\alpha^6$ order contributions may be found in \cite{HFS_H2plus}.

\section{Corrections of order $m\alpha^6$ to the effective Hamiltonian of the fine structure.}

In what follows we use the notation: $\mathbf{R}_1$, $\mathbf{R}_2$, $\mathbf{r}_e$, and $\mathbf{P}_1$, $\mathbf{P}_2$, $\mathbf{p}_e$ are coordinates and impulses of three particles in the center of mass frame, where subscript 1 stands for a helium nucleus and 2 for an antiproton. We also make use $\mathbf{r}_i=\mathbf{r}_e\!-\!\mathbf{R}_i$, $i=1,2$ for coordinates of an electron with respect to one of the nuclei.

According to the NRQED the effective Hamiltonian includes the three new interactions, which contribute to the electron spin-orbit term at the $m\alpha^6$ order. Two are relativistic corrections of order $v^2/c^2$ to the vertex functions for the spin-orbit and Fermi interactions (see \cite{Kin96})
\begin{equation}\label{tree_level_1}
\begin{array}{@{}l}
\displaystyle
\mathcal{V}_1 =
     e^2\left(
        i\frac{3\boldsymbol{\sigma}_e^P[\mathbf{q}\times\mathbf{p}_e]
                                                (p_e'^2+p_e^2)}{32m_e^4}
     \right)
     \;\frac{1}{\mathbf{q}^2}\;(Z_i)
\\[4mm]\displaystyle
\mathcal{V}_2 =
     -e^2
     \left(
        i\frac{[\boldsymbol{\sigma}_e^P\times\mathbf{q}](p_e'^2+p_e^2)}
                                                     {8m_e^3}
     \right)
     \frac{1}{\mathbf{q}^2}
     \left(Z_i\frac{\mathbf{P}_i}{M_i}\right)
\end{array}
\end{equation}
The third is the seagull vertex interaction with one Coulomb and one transverse photon lines:
\begin{equation}\label{seagul_1}
\begin{array}{@{}l}
\displaystyle
\mathcal{V}_3 =
     e^4\frac{\boldsymbol{\sigma}_e^P}{4m_e^2}
     \Biggl[
        \Biggl\{
           \left[
              -\frac{1}{\mathbf{q}_2^2}
              \left(\delta^{ij}\!-\!\frac{q_2^iq_2^j}{\mathbf{q}_2^2}\right)
           \right]
\\[4mm]\hspace{10mm}\displaystyle
           \left(\!-Z_2\frac{\mathbf{P}'_2\!+\!\mathbf{P}_2}{2M_2}\right)
        \Biggr\}
        \left[\frac{i\mathbf{q}_1}{\mathbf{q}_1^2}\right](Z_1)
     \Biggr]+(1\leftrightarrow2)
\end{array}
\end{equation}
the sources may be two different nuclei or may coincide.

To get a complete set of corrections one needs to take into account the second order contribution as well
\begin{equation}\label{2nd_order}
\begin{array}{@{}l}
\displaystyle
\Delta E_A =
   2\alpha^4
   \biggl\langle
      H_B^{}
      \bigg|Q(E_0\!-\!H_0)^{-1}Q\bigg|
      \frac{Z_i(\mathbf{r}_i\!\times\!\mathbf{p}_e)}{2m_e^2r_i^3}
           \mathbf{s}_e
   \biggr\rangle
\\[3mm]\hspace{5mm}\displaystyle
   -2\alpha^4
   \left\langle
      H_B^{}
      \bigg|Q(E_0\!-\!H_0)^{-1}Q\bigg|
      \frac{Z_i(\mathbf{r}_i\!\times\!\mathbf{P}_i)}{m_eM_ir_i^3}
           \mathbf{s}_e
   \right\rangle
\end{array}
\end{equation}
where
\begin{equation}\label{BP}
H_B^{} =
      -\frac{\mathbf{p}_e^4}{8m_e^3}
      +\frac{\pi}{2m_e^2}
         \left[Z_1\delta(\mathbf{r}_1)\!+\!Z_2\delta(\mathbf{r}_2)\right].
\end{equation}

Radiative corrections (form factors of the electron) have been already included into consideration as contributions to the anomalous magnetic moment.

Transforming potentials $\mathcal{V}_i$ to the coordinate space and atomic units one gets:
\begin{equation}\label{V12}
\begin{array}{@{}l}
\displaystyle
\mathcal{V}_1 =
   -\alpha^6c^2\frac{3Z_i}{16m_e^4}
      \left\{
         p_e^2,\frac{1}{r_{i}^3}[\mathbf{r}_{i}\!\times\!\mathbf{p}_e]
      \right\}\mathbf{s}_e,
\\[4mm]\displaystyle
\mathcal{V}_2 =
   \alpha^6c^2\frac{Z_i}{4m_e^3M_i}
      \left\{
         p_e^2,\frac{1}{r_{i}^3}[\mathbf{r}_{i}\!\times\!\mathbf{P}_i]
      \right\}\mathbf{s}_e.
\end{array}
\end{equation}
and
\begin{subequations}
\begin{equation}\label{V3}
\begin{array}{@{}l}
\displaystyle
\mathcal{V}_3=-\alpha^6c^2\frac{Z_1Z_2}{4m_e^2}
    \biggl\{
       \frac{[\mathbf{r}_1\!\times\!\mathbf{P}_2]}{M_2r_1^3r_2}
       +\frac{[\mathbf{r}_2\!\times\!\mathbf{P}_1]}{M_1r_1r_2^3}
\\[4mm]\hspace{7mm}\displaystyle
       -\frac{[\mathbf{r}_1\!\times\!\mathbf{r}_2]}{r_1^3r_2^3}
       \left[
          \frac{(\mathbf{r}_1\mathbf{P}_1)}{M_1}
          -\frac{(\mathbf{r}_2\mathbf{P}_2)}{M_2}
       \right]
    \biggr\}\mathbf{s}_e\,.
\end{array}
\end{equation}
\begin{equation}\label{V4}
\begin{array}{@{}l}
\displaystyle
\mathcal{V}_4=-\alpha^6c^2\frac{1}{4m_e^2}
    \biggl\{
       Z_1^2\frac{[\mathbf{r}_1\!\times\!\mathbf{P}_1]}{M_1r_1^4}
       +Z_2^2\frac{[\mathbf{r}_2\!\times\!\mathbf{P}_2]}{M_2r_2^4}
    \biggr\}\mathbf{s}_e\,.
\end{array}
\end{equation}
\end{subequations}

\section{Variational wave function}

For numerical calculations we use the exponential variational expansion, which
has been discussed in details in \cite{var99}. Namely, the wave function for a state with a total orbital angular momentum $L$ and of a total spatial parity $\pi=(-1)^L$ is expanded as
follows:
\begin{equation}\label{exp_main}
\begin{array}{@{}l}
\displaystyle
\Psi(\mathbf{R},\mathbf{r}_1) =
       \sum_{l_1+l_2=L}
         \mathcal{Y}^{l_1l_2}_{LM}(\hat{\mathbf{R}},\hat{\mathbf{r}}_1)
         G^{L\pi}_{l_1l_2}(R,r_1,r_2),\\[4mm]
\displaystyle\hspace{0mm}
G^{L\pi}_{l_1l_2}(R,r_1,r_2) =
   \sum_{n=1}^N
   \Big\{
      C_n\,\mbox{Re}\bigl[e^{-\alpha_n R-\beta_n r_1-\gamma_n r_2}\bigr]
\\[2mm]\displaystyle\hspace{25mm}
      +D_n\,\mbox{Im}\bigl[e^{-\alpha_n R-\beta_n r_1-\gamma_n r_2}\bigr]
   \Big\}.
\end{array}
\end{equation}
where $\mathbf{R}$ is a position vector of an antiproton and $\mathbf{r}_1$ is a position vector of an electron with respect to a nucleus; parameters $\alpha_n$ are complex, and $\beta_n$, $\gamma_n$ are real, they are generated in a pseudorandom way,
\begin{equation}\label{generators}
\begin{array}{@{}l}
\displaystyle
\mathrm{Re}(\alpha_n) = \left[\left\lfloor{\textstyle\frac{1}{2}}n(n+1)\sqrt{p_\alpha}
                \right\rfloor(A_2-A_1)+A_1\right],
\\[3mm]\displaystyle
\mathrm{Im}(\alpha_n) = \left[\left\lfloor{\textstyle\frac{1}{2}}n(n+1)\sqrt{p'_\alpha}
                \right\rfloor(A'_2-A'_1)+A'_1\right],
\\[3mm]\displaystyle
\beta_n = \left[\left\lfloor{\textstyle\frac{1}{2}}n(n+1)\sqrt{p_\beta}
                \right\rfloor(B_2-B_1)+B_1\right],
\\[3mm]\displaystyle
\gamma_n = \left[\left\lfloor{\textstyle\frac{1}{2}}n(n+1)\sqrt{p_\gamma}
                \right\rfloor(C_2-C_1)+C_1\right].
\end{array}
\end{equation}
Here $\lfloor x\rfloor$ denotes fractional part of $x$, and $p_\alpha$, $p_\beta$ or $p_\gamma$ are some prime numbers and $A_i$, $B_i$, $C_i$ are variational parameters.

For the initial wave function of the bound state we use the triple basis set with the total number of terms $N = 2200$ in expansion (\ref{exp_main}) and full optimization of variational parameters. That yields the non-relativistic energy for this state
\[
E_{nr}(37, 35) = -2.899\,282\,183\,295\,31(1) \mbox{ au}
\]
The CODATA06 recommended values \cite{CODATA06} have been adopted for calculations: $m_{\bar{p}}=m_p=1836.152\,672\,47(80)m_e$, $m_\alpha = 7294.299\,5365(31) m_e$ and $R_\infty c =3.289\,841\,960\,361(22) \times 10^6$ MHz.

For the intermediate states of the second order iteration the similar variational expansion (\ref{exp_main}) with various basis lengths $N=520\!\div\!960$ has been used.

\begin{table}[t]
\begin{center}
\begin{tabular}{r@{\hspace{4mm}}c@{\hspace{4mm}}c@{\hspace{4mm}}c@{\hspace{4mm}}c}
\hline\hline
 set~~~ & $[A_1,A_2]$ & $[A_1',A_2']$ & $[B_1,B_2]$ & $[C_1,C_2]$ \\
\hline
1-st set & [66.6, 87.6] & [0.4, 5.2] & [0.00, 2.05] & [0.00, 0.87] \\
2-nd set & [66.0, 75.4] & [0.0, 5.4] & [0.94, 5.70] & [0.00, 1.94] \\
 3-d set & [66.0, 75.4] & [0.0, 5.4] & [5.00, 80.0] & [0.00, 0.10] \\
4-th set & [66.0, 75.4] & [0.0, 5.4] & [0.00, 0.20] & [2.00, 70.0] \\
5-th set & [66.0, 75.4] & [0.0, 5.4] & [90., 1000.] & [0.00, 0.10] \\
6-th set & [66.0, 75.4] & [0.0, 5.4] & [0.00, 0.10] & [80.0, 800.] \\
7-th set & [66.0, 75.4] & [0.0, 5.4] & [$10^3$, $10^4$] & [0.00, 0.10] \\
8-th set & [66.0, 75.4] & [0.0, 5.4] & [0.00, 0.10] & [800., $10^4$] \\
\hline\hline
\end{tabular}
\end{center}
\caption{Variational parameters for eight basis sets used in the second order contribution calculations.}\label{var_par}
\end{table}

\section{Reduce a singularity in the second order contribution}

The $H_B^{}$ operator in the second order term (\ref{2nd_order}) is too singular. It requires careful consideration because intermediate states should include functions with asymptotic behaviour at small distances like $\sim\!1/r_1$ (or $1/r_2$). The usual regular trial functions would result in a very slow convergence of $\Delta E_A$.

In order to smooth the perturbation and to reduce the singularity of the intermediate wave function we may use transformation
\begin{equation}\label{HPP}
{H'_{B\!}}=H_B^{}-(E_0-H_0)U-U(E_0-H_0)
\end{equation}
The delta-function singularity in $|H_B^{}\Psi_0\rangle$ has the following structure
\begin{equation}\label{HP}
\begin{array}{@{}l}
\displaystyle
H_B\Psi_0 =
   -\frac{1}{m_e^2}\biggl[
      Z_1\left(\frac{\mu_1}{m_e}-\frac{1}{2}\right)\pi\delta(\mathbf{r}_1)
\\[3mm]\hspace{16mm}\displaystyle
      +Z_2\left(\frac{\mu_2}{m_e}-\frac{1}{2}\right)\pi\delta(\mathbf{r}_2)
   \biggr]\Psi_0
   +\cdots,
\end{array}
\end{equation}
where $1/\mu_i = 1/m_e+1/M_i$.

It is natural to take $U$ in the form $U=c_1/r_1+c_2/r_2$. The coefficients $c_i$ may be obtained by substituting $U$ into the initial Schr\"odinger equation
\[
\begin{array}{@{}l}
\displaystyle
(E_0-H_0)\left(\frac{c_1}{r_1}+\frac{c_2}{r_2}\right)
\\[3mm]\hspace{10mm}\displaystyle
=
   -\frac{2c_1}{\mu_1}\pi\delta(\mathbf{r}_1)
   -\frac{2c_2}{\mu_2}\pi\delta(\mathbf{r}_2)+\dots
\end{array}
\]
then comparing the latter expression with Eq.~(\ref{HP}) one gets:
\begin{equation}
\begin{array}{@{}l}
\displaystyle
c_1 = \frac{\mu_1(2\mu_1\!-\!m_e)}{4m_e^3}\>Z_1,
\\[3mm]\displaystyle\mbox{\vrule width 0pt depth 19pt}
c_2 = \frac{\mu_2(2\mu_2\!-\!m_e)}{4m_e^3}\>Z_2.
\end{array}
\end{equation}
Thus the second order term may be rewritten as follows
\begin{equation}\label{2nd_order_m}
\begin{array}{@{}l}
\left\langle
   H_B^{}|Q(E_0-H_0)^{-1}Q|H_{SO}^{}
\right\rangle
\\[2mm]\hspace{6mm}
=
\left\langle
   H_B^{'}|Q(E_0-H_0)^{-1}Q|H_{SO}^{}
\right\rangle
\\[2.5mm]\hspace{15mm}
+\left\langle
   UH_{SO}^{}
\right\rangle
-\left\langle U \right\rangle \left\langle H_{SO}^{} \right\rangle).
\end{array}
\end{equation}
Matrix elements of $H_B'$ may be obtained directly from Eq.~(\ref{HPP}). Additional term to the effective Hamiltonian is expressed
\begin{equation}\label{H_m}
\begin{array}{@{}l}
\displaystyle
\left\langle H_m^{(6)} \right\rangle =
   \left\langle
      UH_{SO}^{}
   \right\rangle
   -\left\langle U \right\rangle \left\langle H_{SO}^{} \right\rangle
\\[3mm]\hspace{4mm}\displaystyle
=
   \left\langle
      \left(\frac{c_1}{r_1}+\frac{c_2}{r_2}\right) H_{SO}^{}
   \right\rangle
   -\left\langle \frac{c_1}{r_1}+\frac{c_2}{r_2} \right\rangle
      \left\langle H_{SO}^{} \right\rangle.
\end{array}
\end{equation}

For the numerical evaluation of the second order term from Eq.~(\ref{2nd_order_m}) we use the eight basis sets, where the first two approximate the regular part of the intermediate solution, and the remaining six sets with growing exponents are introduced to reproduce behaviour of the type $\ln(r_1)$ (or $\ln(r_2)$) at small values of $r_1$ or $r_2$. The particular variational parameters used are presented in Table \ref{var_par}.

\section{Numerical calculations}

\begin{table*}[t]
\begin{center}
\begin{tabular}{r@{\hspace{2mm}}r@{\hspace{2mm}}r@{\hspace{7mm}}r@{\hspace{7mm}}r@{\hspace{7mm}}r@{\hspace{5mm}}r@{\hspace{5mm}}r}
\hline\hline
\vrule width 0pt height 16.5pt depth 11 pt
$n_1$ & $n_2$ & $n_3$
& $\displaystyle\frac{[\mathbf{r}_1\times\mathbf{p}_e]}{r_1^3}$
& $\displaystyle\frac{[\mathbf{r}_1\times\mathbf{P}_1]}{r_1^3}$
& $\displaystyle\frac{[\mathbf{r}_2\times\mathbf{p}_e]}{r_2^3}$
& $\displaystyle\frac{[\mathbf{r}_2\times\mathbf{P}_2]}{r_2^3}$ \\
\hline
20 & 20 &  0 & 0.2883781 & 680.0299 & 0.5135404 & $-$1204.595 \\
20 & 20 & 20 & 0.2411901 & 624.7925 & 0.4875501 & $-$1177.070 \\
40 & 20 & 20 & 0.2242844 & 573.8206 & 0.4899458 & $-$1192.019 \\
60 & 20 & 20 & 0.2536075 & 633.2949 & 0.5022857 & $-$1194.613 \\
60 & 40 & 20 & 0.2232982 & 526.5736 & 0.4881029 & $-$1187.319 \\
60 & 40 & 40 & 0.2847268 & 714.1212 & 0.4849738 & $-$1160.058 \\
80 & 40 & 40 & 0.2812069 & 716.9230 & 0.4696484 & $-$1159.465 \\
\hline\hline
\end{tabular}
\end{center}
\caption{Convergence of the second order contribution matrix elements for the spin-orbit interaction.}\label{tab:conv}
\end{table*}

Matrix elements in Eqs.~(\ref{V12})-(\ref{V3}) and (\ref{2nd_order_m})-(\ref{H_m}) for the basis functions (\ref{exp_main}) of the exponential variational expansion were evaluated analytically using the recurrences derived in \cite{KorJPB02} with some modifications, which allowed to improve stability. The generating functions $\Gamma_{-4,0,0}(\alpha,\beta,\gamma)$, $\Gamma_{-3,-1,0}(\alpha,\beta,\gamma)$ were taken from \cite{Harris}. What corresponds to the cut-off regularization of the integrals at $r_{1,2}=\rho$, where $\rho\ll1$.

For all vector operators the reduced matrix element is assumed. Here we present numerical values of some the most complicate operators.
\begin{widetext}
\begin{equation}
\begin{array}{@{}l@{\hspace{25mm}}l}
\displaystyle
\left\langle
   p_e^2\;\frac{[\mathbf{r}_1\times\mathbf{p}_e]}{r_1^3}
\right\rangle = 1.5575929
&\displaystyle
\left\langle
   p_e^2\;\frac{[\mathbf{r}_2\times\mathbf{p}_e]}{r_2^3}
\right\rangle = 2.2459243
\\[4mm]\displaystyle
\left\langle
   p_e^2\;\frac{[\mathbf{r}_1\times\mathbf{P}_1]}{r_1^3}
\right\rangle = 3272.7020
&\displaystyle
\left\langle
   p_e^2\;\frac{[\mathbf{r}_2\times\mathbf{P}_2]}{r_2^3}
\right\rangle = -3080.3879
\end{array}
\end{equation}
\begin{equation}
\begin{array}{@{}l@{\hspace{25mm}}l}
\displaystyle
\left\langle
   \frac{1}{r_2}\;\frac{[\mathbf{r}_1\times\mathbf{P}_2]}{r_1^3}
\right\rangle = 205.83272
&\displaystyle
\left\langle
   \frac{1}{r_1}\;\frac{[\mathbf{r}_2\times\mathbf{P}_1]}{r_2^3}
\right\rangle = 1391.5321
\\[4mm]\displaystyle
\left\langle
   \frac{[\mathbf{r}_1\times\mathbf{r}_2]}{r_1^3r_2^3}(\mathbf{r}_1\mathbf{P}_1)
\right\rangle = 551.65420
&\displaystyle
\left\langle
   \frac{[\mathbf{r}_1\times\mathbf{r}_2]}{r_1^3r_2^3}(\mathbf{r}_2\mathbf{P}_2)
\right\rangle = 551.43328
\end{array}
\end{equation}
\begin{equation}\label{2nd_order_mm}
\begin{array}{@{}l@{\hspace{10mm}}l}
\displaystyle
\left\langle
   H_B'\Bigl|Q(E_0\!-\!H_0)^{-1}Q\Bigr|
   \frac{[\mathbf{r}_1\times\mathbf{p}_e]}{r_1^3}
\right\rangle = 0.2812
&\displaystyle
\left\langle
   H_B'\Bigl|Q(E_0\!-\!H_0)^{-1}Q\Bigr|
   \frac{[\mathbf{r}_2\times\mathbf{p}_e]}{r_2^3}
\right\rangle = 0.4696
\\[4mm]\displaystyle
\left\langle
   H_B'\Bigl|Q(E_0\!-\!H_0)^{-1}Q\Bigr|
   \frac{[\mathbf{r}_1\times\mathbf{P}_1]}{r_1^3}
\right\rangle = 717.
&\displaystyle
\left\langle
   H_B'\Bigl|Q(E_0\!-\!H_0)^{-1}Q\Bigr|
   \frac{[\mathbf{r}_2\times\mathbf{P}_2]}{r_2^3}
\right\rangle = -1160.
\end{array}
\end{equation}
\end{widetext}

It is worthy to say that the second order iteration, even after reduction of the singularity, still reveals slow convergence. In Table \ref{tab:conv} we present results of numerical calculations for various sets of basis functions. The results depend very little on increase of the 1st and 2nd basis sets (see Table \ref{var_par}), which represent regular behaviour. The following notation has been used in Table \ref{tab:conv}: $n_1$ is a number of basis functions for 3d and 4th sets, $n_2$ is for 5th and 6th sets, etc. As is seen from the Table, no more than two digits may be accepted with confidence as convergent. However, the increase of the basis sets leads to numerical instability, which we attribute to very large angular momentum of the state ($L=35$) what makes the recursion used for analytic evaluation of the matrix elements to be too long and unstable for large exponents. The octuple precision has been used in these calculations and still it was not enough to provide necessary stability.

\section{Results and discussion}

Summing up the contributions from Eq.~(\ref{2nd_order}), (\ref{V12}), and (\ref{V3})-(\ref{V4}) we obtain the $m\alpha^6$ order contribution to the electron spin-orbit interaction:
\begin{equation}\label{D_E1}
\Delta E_1 = -0.000030(4)\cdot10^{-7} \mbox{ au}
\end{equation}
Thus a new value for the $E_1$ coefficient would be
\begin{equation}\label{E1}
E_1 = -0.552\,563(4)\cdot10^{-7} \mbox{ au}
\end{equation}
where the uncertainty is primarily due to slow convergence of the second order iteration.

Using this new value for the $E_1$ coefficient and keeping $E_2$--$E_4$ as in \cite{JPB01} one may solve the effective Hamiltonian (\ref{eq:eff_H}) and get updated theoretical values for transition frequencies:
\begin{equation}
\begin{array}{@{}l}
\tau_+ = 12897.0(1)(3) \mbox{ MHz} \\
\tau_- = 12924.9(1)(3) \mbox{ MHz} \\
\Delta\tau = 27.897(0)(3) \mbox{ MHz}
\end{array}
\end{equation}
The first error indicates the numerical uncertainty of present calculations, while the second one is an estimate of the theoretical uncertainty due to yet uncalculated higher order terms. As it should be $\Delta\tau$ does not change its value comparing with previous calculations \cite{JPB01}. The values of $\tau_+$ and $\tau_-$ leap over the experimental result and become to overestimate ones if we take the numerical error as a measure of uncertainty. Still in theory we need to include into consideration effects of the next order in $\alpha$, which contain terms of order $(\alpha^3\ln{\alpha})E_1$ and are of the magnitude of the discrepancy.

It is worthy to note here that the obtained value of $\Delta E_1$ is unexpectedly small. That explains rather good agreement of the experiment with the results of the Breit-Pauli approximation.

In order to get the improved value for $\Delta\tau$ one needs to perform a complete calculation of all the contributions of order $m\alpha^6(m/M)$, which provides corrections to the remaining coefficients, $E_2-E_4$, in the effective HFS Hamiltonian (\ref{eq:eff_H}). This work is in progress now.

\section{Acknowledgments}

The support of the Russian Foundation for Basic Research under Grants No. 08-02-00341 and 09-02-91000-ASF is gratefully acknowledged.


\begin{thebibliography}{99}
\bibitem{PRA98} D.~Bakalov and V.I.~Korobov, Phys.\ Rev.~A \textbf{57}, 1662 (1998).
\bibitem{Kino01} N.~Yamanaka, Y.~Kino, H.~Kudo, and M.~Kamimura, Phys.\ Rev.~A
  \textbf{63}, 012518 (2001).
\bibitem{JPB01} V.I.~Korobov and D.~Bakalov, J.~Phys.~B: At.\ Mol.\ Opt.\ Phys.
  \textbf{34}, L519 (2001)
\bibitem{Kino03} Y.~Kino, N.~Yamanaka, M.~Kamimura, and H.~Kudo, Hyperfine Interactions
  \textbf{146/147}, 331 (2003).
\bibitem{Wid02} E.~Widmann, J.~Eades, T.~Ishikawa, J.~Sakaguchi, T.~Tasaki, H.~Yamaguchi,
  R.S.~Hayano, M.~Hori, H.A.~Torii, B.~Juh\'asz, D.~Horv\'ath, and T.~Yamazaki,
  Phys.\ Rev.\ Lett. \textbf{89}, 243402 (2002).
\bibitem{Wid08} T.~Pask, D.~Barna, A.~Dax, R.S.~Hayano, M.~Hori,
  D.~Horv\'ath, B.~Juh\'asz, C.~Malbrunot, J.~Marton, N.~Ono, K.~Suzuki,
  J.~Zmeskal, and E.~Widmann, J.~Phys.~B \textbf{41}, 081008 (2008).
\bibitem{PLB09} T.~Pask, D.~Barna, A.~Dax, R.S.~Hayano, M.~Hori, D.~Horv\'ath, S.~Friedreich,
  B.~Juh\'asz, O. Massiczek, N. Ono, A. S\'ot\'er, and E.~Widmann, Phys.\ Lett.~B
  \textbf{678}, 55 (2009).
\bibitem{Kin96} T.~Kinoshita and M.~Nio, Phys.\ Rev.~D \textbf{53}, 4909
  (1996).
\bibitem{HFS_H2plus} V.I.~Korobov, L.~Hilico, and J.-Ph.~Karr, Phys.\ Rev.~A \textbf{79},
  012501 (2009).
\bibitem{var99} V.I.~Korobov, D.~Bakalov, and H.J.~Monkhorst,
  Phys.\ Rev.~A, {\bf 59}, R919 (1999).
\bibitem{CODATA06} P.J.~Mohr, B.N.~Taylor, and D.B.~Newell, Rev.\ Mod.\ Phys.\ \textbf{80},
  633 (2008).
\bibitem{KorJPB02} V.I.~Korobov, J.~Phys.~B: At.\ Mol.\ Opt.\ Phys.\ \textbf{35}, 1959
  (2002).
\bibitem{Harris} F.E.~Harris, A.M.~Frolov, and V.H.~Smith, Jr., J.~Chem.\ Phys.
  \textbf{121}, 6323 (2004).
\end{thebibliography}
\end{document}